\documentclass[11pt,prd,preprintnumbers,nofootinbib,superscriptaddress,notitlepage]{revtex4-1}

\usepackage{graphics}
\usepackage{bm}
\usepackage{cancel}
\usepackage{color}
\usepackage{xcolor}
\usepackage{multirow}
\usepackage{slashed}
\usepackage{array}
\usepackage{amssymb}
\usepackage{bbold}
\usepackage{graphicx}
\usepackage{mathrsfs}
\usepackage{diagbox}
\usepackage{amsmath}
\usepackage{float}
\usepackage{extarrows}
\usepackage{booktabs}
\usepackage[colorlinks,citecolor=blue,anchorcolor=red,menucolor=red,linkcolor=red,filecolor=red,runcolor=red,urlcolor=red,frenchlinks=red]{hyperref}
\usepackage{subfigure}
\usepackage{ulem}

\usepackage{bm}

\makeatletter
\newcommand{\thickhline}{\noalign {\ifnum 0=`}\fi \hrule height 1pt\futurelet \reserved@a \@xhline}
\newcolumntype{"}{@{\hskip\tabcolsep\vrule width 1pt\hskip\tabcolsep}}
\makeatother

\begin{document}
\title{Study $\gamma\gamma \to \tau^+\tau^-$ process including $\tau^+ \tau^-$ spin information in Pb-Pb ultraperipheral collision and at Lepton collider}

\author{Peng-Cheng Lu}
\email{pclu@sdu.edu.cn}
\affiliation{School of Physics, Shandong University, Jinan, Shandong 250100, China}

\author{Zong-Guo Si}
\email{zgsi@sdu.edu.cn}
\affiliation{School of Physics, Shandong University, Jinan, Shandong 250100, China}

\author{Han Zhang}
\email{han.zhang@mail.sdu.edu.cn}
\affiliation{School of Physics, Shandong University, Jinan, Shandong 250100, China}

\author{Xin-Yi Zhang}
\email{xinyizhang@mail.sdu.edu.cn}
\affiliation{School of Physics, Shandong University, Jinan, Shandong 250100, China}
\begin{abstract}

We study  the  $\gamma\gamma \to \tau^+\tau^-$ process including full $\tau^+ \tau^-$ spin information in Pb--Pb ultraperipheral collision  and at lepton colliders.
We present the predictions for the corresponding  cross sections and spin correlations at NLO electroweak precision, 
and find that the NLO electroweak contributions are numerically small for the observables considered.
Additionally, we use the spin correlations obtained in this paper to analyze the quantum entanglement in the $\tau^+\tau^-$ system of the  $\gamma\gamma \to \tau^+\tau^-$ process. Our results show that there is a genuine entangled configuration near the $\tau^+\tau^-$ invariant mass threshold. 
This work is helpful for studying the $\tau$-pair production induced by photon-photon collision at high energy colliders.

\end{abstract}
\maketitle

\section{Introduction}
\label{sec:introduction}
The heaviest charged lepton in the Standard Model (SM) is the $\tau$ lepton, with a mass of\(\sim 1.777\;\mathrm{GeV}/c^2\) and a lifetime of \(\sim 2.9 \times 10^{-13}\;\mathrm{s}\)~\cite{websitpdg}, exhibiting both leptonic and hadronic decay modes with precisely measured branching ratios~\cite{ALEPH:2005qgp,ParticleDataGroup:2024cfk}, making it a powerful probe to study electroweak (EW) couplings and to search for physics beyond the SM.
Since $\tau$ lepton discovery, its  properties and interactions have been studied extensively not only at lepton colliders (LC), such as LEP, BEPC, and (Super)KEKB~\cite{ALEPH:2001uca, DELPHI:2003zcz,Belle:2006qqw,Zhang:2018gol}, but also at hadron collider\cite{ATLAS:2018ynr,CMS:2024qjo, ATLAS:2022akr}.
More recently, both ATLAS and CMS reported the observation of photon induced $\tau$-pair production process in ultraperipheral lead-lead collisions (UPC) at LHC, which facilitated tighter constraints on beyond SM physics  and more precise understanding of  $\tau$ lepton's properties~\cite{ATLAS:2022ryk,CMS:2022arf}.
Although the $\tau$ lepton decays rapidly, its polarization effects are not washed out. 
They can be reconstructed via the angular and energy distributions of its decay final states, known as tau spin-analyzers~\cite{Tsai:1971vv,ALEPH:2001uca}. 
Since these spin-polarization and spin-correlation phenomena reflect in detail the interactions involved in $\tau$ lepton production and decay,  the study of them plays an important role.

As far as theoretical predictions are concerned, spin effects in $\tau$-pair production and decay are typically described using the spin density matrix formalism, which allows for the extraction of polarization and spin correlation observables in various  processes. 
The spin density matrix can be reconstructed using either decay method~\cite{Kuhn:1990ad} or kinematic method \cite{Cheng:2024rxi}. 
The spin-dependent observables are sensitive to, such as anomalous couplings~\cite{Bernabeu:2007rr}, CP-violating effects~\cite{Chen:2018cxt} and other new physics signal~\cite{Bar-Shalom:1998rqq,Bullock:1992yt}, which have been studied in multiple processes, in $e^+e^- \to \tau^+ \tau^-$~\cite{Bodrov:2024wrw, Lu:2025heu}, in the process including resonance production such as $\Upsilon \to \tau^+ \tau^-$ and $J/\psi \to \tau^+ \tau^-$ at BEPC and Super Tau-Charm Facility~\cite{Han:2025ewp,Sun:2024vcd}, as well as in hadronic collisions via $pp \to  \tau^+ \tau^-$ at LHC~\cite{CMS:2024qjo}. 
Furthermore, recent studies explored the quantum entanglement of the spin system and proposed observables that could be used to test Bell inequalities~\cite{PhysRevLett.23.880}  and the concurrence~\cite{Wootters:1997id} which provide a measure of bipartite quantum correlations at colliders.
The first system that is used to study quantum entanglement at the LHC is the $t\bar{t}$ system~\cite{ATLAS:2023fsd,CMS:2024pts}. 
This was extended to $\tau^+\tau^-$ system and others~\cite{Ashby-Pickering:2022umy,Wu:2024ovc}. Ref.\cite{han2025entanglementbellnonlocalitytau,Fabbrichesi:2024wcd,Ehataht:2023zzt} and \cite{Zhang:2025mmm} demonstrated that $\tau^+\tau^-$ systems exhibit entanglement over a wide range of energies in the future $e^+e^-$ colliders and at LHC.
In this paper we extend the analysis to the process $\gamma\gamma \to \tau^+\tau^-$ 
both in UPC collision at LHC and at future LC, where photon induced $\tau$-pair production can be studied with high precision. 

The studies of photon-photon ($\gamma\gamma$) process have been carried out in proton-proton, proton-nucleus, and nucleus-nucleus UPC and at LC via Compton back scattering of laser photons off high-energy electron beams.
In UPC, the Lorentz-boosted electromagnetic fields of heavy nuclei serve as intense sources of quasi-real photons, with fluxes scaling as $Z^4$ in  nucleus-nucleus collisions, where $Z$ represents the nuclear charge.
Such processes are described by the equivalent photon approximation (EPA)~\cite{vonWeizsacker:1934nji,Williams:1934ad}, offering a clean environment with very low hadronic backgrounds and pile-up for multi-TeV $\gamma\gamma$ studies~\cite{Dyndal:2023sts}.
At LC, the same EPA framework—often referred to as the Weizsäcker–Williams approximation (WWA)~\cite{Frixione:1993yw}—is applied to initial-state photons. These machines combine high luminosities and tunable energies with clean final states, enabling precise reconstruction of $\tau$ decays and detailed polarization studies. 
Future facilities such as CEPC~\cite{CEPCStudyGroup:2018ghi} and CLIC~\cite{CLIC:2018fvx} will further extend this reach with higher luminosities and energies.
Muon colliders~\cite{Delahaye:2019omf,MuonCollider:2022ded} provide an additional frontier, combining multi-TeV energies with reduced synchrotron radiation. 
This high level of experimental precision necessitates equally precise theoretical predictions. 
In this context, the next-to-leading order (NLO) quantum electrodynamics corrections for $\gamma\gamma \to \tau^+ \tau^-$ in UPCs were obtained in Ref.~\cite{Shao:2024dmk}. 
Following this, we studied the NLO EW corrections for the unpolarized $\tau$-pair production process~\cite{Jiang:2024dhf}. 
Further developments include the creation of automated software for NLO-accurate predictions of $\gamma\gamma \to \tau^+ \tau^-$ in UPCs~\cite{Shao:2025bma}, 
as well as the  computation of the leptonic decays of spin-correlated $\tau$ leptons, including NLO EW corrections in Pb-Pb UPC~\cite{Dittmaier:2025ikh}.
Parallel investigations have explored the $\gamma\gamma\to\tau^+\tau^-$ process as a probe for new physics effects at LC~\cite{Billur:2013rva,Rajaraman:2018uyb, Koksal:2018xyi, Wang:2024bfc}.
Building on our previous study of the NLO EW corrections to unpolarized $\tau$-pair production induced by $\gamma\gamma$ collisions~\cite{Jiang:2024dhf}, this paper extends the analysis by incorporating the full $\tau^+ \tau^-$ spin information.
The spin formalism we used is based on Ref.~\cite{Bernreuther_2015}, allowing the extraction of polarization and correlation observables relevant for experimental analyses. Our results provide a precision SM baseline for future studies.
  
The structure of this paper is as follows. 
In section \ref{sec2}, we introduce the theoretical framework for $\gamma\gamma \to \tau^+\tau^-$ in Pb-Pb UPC and at LC, reviewing the spin density matrix of $\tau$-pair production and the associated spin correlations.
In Section \ref{sec4}, we present the numerical results for the cross section as well as spin correlation effects corresponding to different colliders setups. We also discuss the prospects of probing quantum entanglement through this process.
Section \ref{sec5} is reserved for the summary.

\section{theoretical framework AND SPIN-CORRELATIONs EFFECT}\label{sec2}
\begin{figure}[htbp]
	\begin{center}
		\subfigure[]{\label{NLO Invariant mass distribution}
			\includegraphics[width=0.36\textwidth]{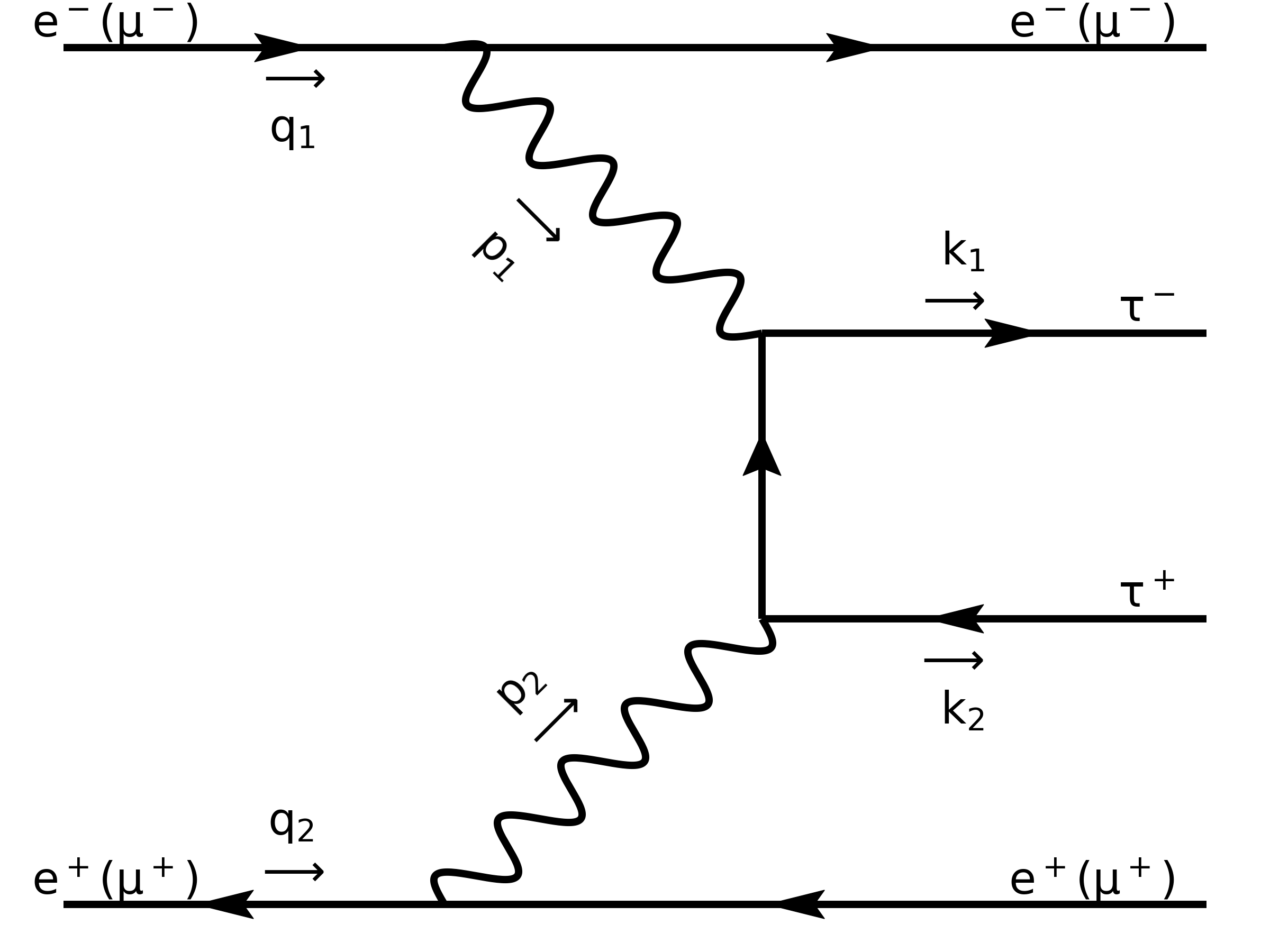} }
		\hspace{-0.3cm}~
		\subfigure[]{\label{NLO pt distribution}
			\includegraphics[width=0.36\textwidth]{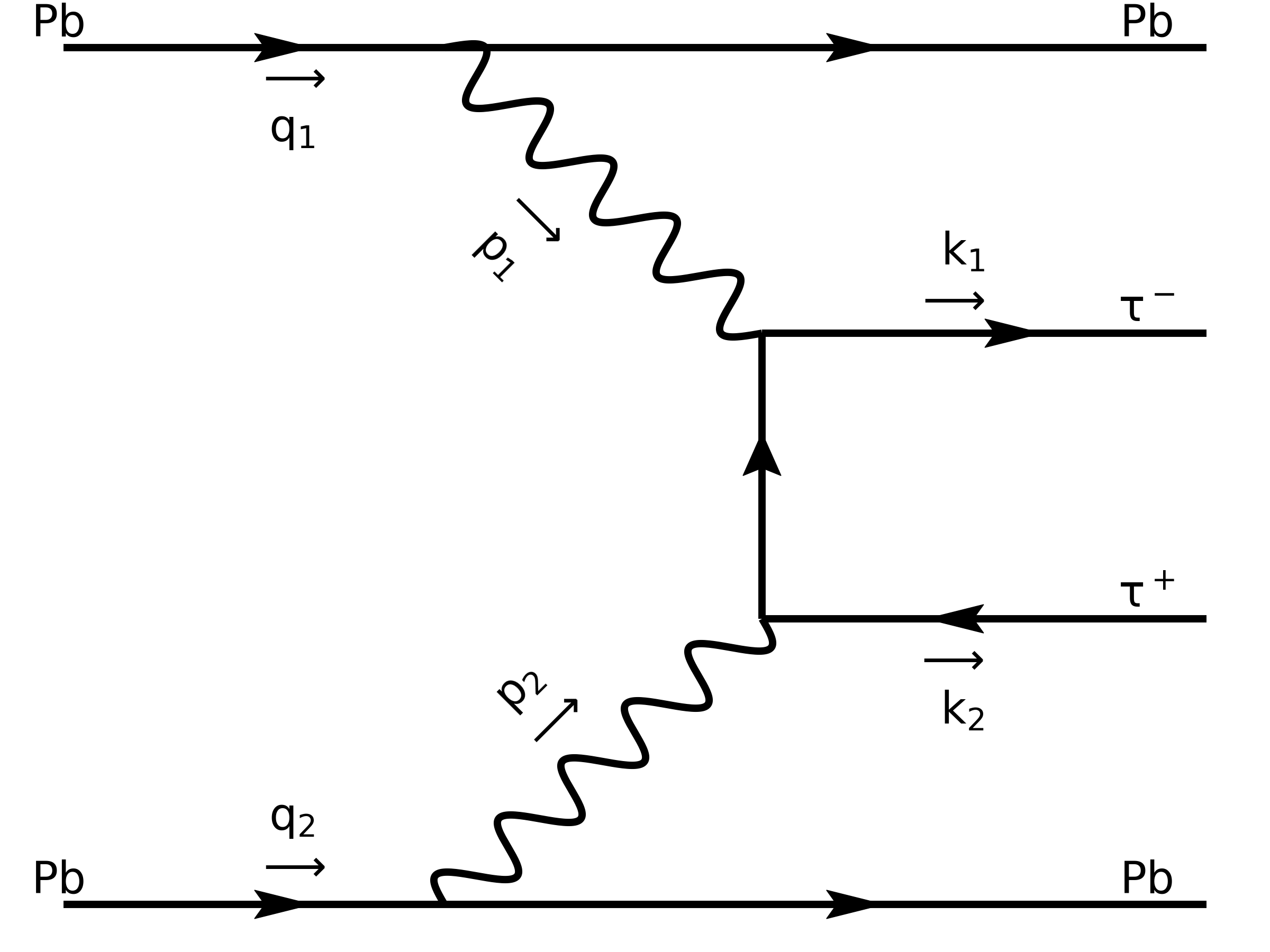} }
		\caption{$t$-channel $\tau^+\tau^-$ production via photon fusion at LC (left) and in Pb-Pb UPC (right). The corresponding $u$-channel contributions are these diagrams involving $\tau$-pair exchange.}
	\label{pb-pb}
	\end{center}
\end{figure}

This section presents the calculation framework for photon induced $\tau$ lepton pair production at LC and in Pb-Pb UPC,  as illustrated in Fig. \ref{pb-pb}, incorporating NLO EW corrections and the effects of $\tau$ lepton spin.
The total cross section for this process can be described as the convolution of the two-photon  flux luminosity with the partonic cross section of the subprocess $\gamma\gamma \to \tau^+ \tau^-$, denoted as $\hat{\sigma}(\gamma \gamma \to \tau^+ \tau^-)$. It is given by:
\begin{equation}
\sigma_{\mathrm{UPC/LC}} = \int \mathrm{d}x_{1} \mathrm{d}x_{2} n(x_{1}) n(x_{2}) \hat{\sigma}(\gamma \gamma \to \tau^{+}\tau^{-}).
\label{eq:sigma_total}
\end{equation}

For UPC events, we adopt the equivalent photon approach based on the EDFF~\cite{Cahn:1990jk}.
In this framework, the photon flux is constructed directly from the coherently emitted photons of the heavy ion.
The coherent photon emission is dominated by small photon virtualities, with a typical scale $Q^2 \sim 1/R_A^2$.
For a heavy nucleus with mass number $A>16$ and a charge radius $R_A \simeq 1.2\,A^{1/3}\,\mathrm{fm}$, this corresponds to $Q^2 \lesssim 4\times10^{-3}\,\mathrm{GeV}^2$~\cite{Shao:2022cly}.
As a consequence, the equivalent photon approximation provides an adequate description of the radiated photons in photon-induced processes in UPC, and the photon distribution function  (PDF) can be formulated as:
\begin{equation}\label{n(x)}
\begin{aligned}
n(x_i)=\frac{2Z^{2} \alpha}{x_i \pi } \left \{ \bar{x}_iK_{0}(\bar{x}_i ) K_{1}(\bar{x}_i)-\frac{\bar{x}_i^{2}}{2} [K_{1}^{2}(\bar{x}_i)-K_{0}^{2}(\bar{x}_i) ]  \right \}.
\end{aligned}
\end{equation}
Here, $n(x_i)$ is the effective photon spectrum integrated over impact parameter $b$ \cite{Jackson:1998nia,Shao:2022cly}, charge number $Z=82$  for Pb, $x_{i}=E_{i}/E_{beam}$ is the ratio of the energy $E_{i}$ of the emitted photons from ion $i$ to the beam energy $E_{beam}$, $\bar{x}_i =x_{i} m_{N} b_{min} $,  $m_{N}$ is the mass of the nucleon $m_{N}=0.9315$ GeV, the minimum of impact parameter $b_{min}$ is set to be the nuclear radius $b_{min} =R_{A} \approx 1.2A^{\frac{1}{3} } \mathrm{fm} = 6.09A^{\frac{1}{3} }$ GeV$^{-1}$ with $A=208$ for Pb, and $K_0,~K_1$ are the modified Bessel functions of the second kind of zero and first order, respectively.
In the UPC, photons emitted by heavy ions are coherently radiated, which imposes a limit on the maximum of center-of-mass (CM) energy of two photons, $\sqrt{s_{\gamma\gamma}^{max} } =\frac{\gamma_L}{R_A}$
where $\gamma_L$ is the Lorentz factor $\gamma_L = E_{beam}/m_N$ \cite{Baur:2001jj}.

At LC,  the initial photons are emitted from the incoming leptons through bremsstrahlung, and their energy distributions can be described using the WWA~\cite{Budnev:1975poe}. Within this framework, the photon luminosity can be derived from the convolution of the PDF associated with the emitting leptons.
The equivalent PDF  of an electron or muon under the WWA is given by \cite{Budnev:1975poe}
\begin{equation}
n(x_i) = \frac{\alpha}{2\pi} \left[ \frac{1 + (1 - x_i)^2}{x_i} \ln\left( \frac{Q^2_{\max}}{Q^2_{\min}} \right) + \frac{2m_f^2 x_i}{Q^2_{\max}} - \frac{2m_f^2 x_i}{Q^2_{\min}} \right],
\label{eq:wwa}
\end{equation}
where $x_i = E_\gamma / E_f$ is the fraction of the energy carried by the photon, $\alpha$ is the fine-structure constant, and $m_f$ is the  mass of incoming lepton, 
$m_f = m_e (m_{\mu})$ for  $e^+ e^-$ (Muon) collider. 
The photon virtuality bounds $Q^2_{\min}$ and $Q^2_{\max}$ are determined by
\begin{equation}
\begin{aligned}
&Q^2_{\min} = \frac{m_f^2 x_i^2}{1 - x_i},\  &Q^2_{\max} = \frac{(E_f \theta_c)^2}{1 - x_i} + Q^2_{\min},
\end{aligned}
\end{equation}
where $\theta_c$ denotes the maximal electron scattering angle allowed by the detector acceptance, with a typical value of $\theta_c = 32$ mrad used at $e^+e^-$ colliders~\cite{Klasen:2001cu} and Muon collider.
Given the clean initial state and high-precision environment of LC, the $f^+ f^- \stackrel{\gamma \gamma}{\longrightarrow} f^+ f^- \tau^+ \tau^-$ process, with $f^\pm = e^\pm$ or $\mu^\pm$, provides a promising channel for studying the properties of the tau lepton.

We consider NLO EW effects and spin correlations in the subprocess
\begin{equation}\label{eq:ggtautau}
 \gamma(p_1,\lambda_1) + \gamma(p_2,\lambda_2) \;\to\;
 \tau^-(k_1,s_1) + \tau^+(k_2,s_2) \, ,
\end{equation}
where $p_{1,2}$ and $k_{1,2}$ denote the four-momenta of the incoming photons and outgoing $\tau$ leptons, respectively. $\lambda_{1,2}=\pm1$ are the photon helicities. The spin states of the $\tau^-$ and $\tau^+$ are represented by the spin four-vectors $s_{1,2}$, respectively, and $k_1 \cdot s_1 = k_2 \cdot s_2 = 0$.
In the rest frame of $\tau^-$ ($\tau^+$), the spin is described by a unit three-vectors $\vec{s}_1$ ($\vec{s}_2$).  With the aid of Lorentz boost transformation, we have the components of the Pauli-Lubanski 4-vector $s^\mu_1$ and $s^\mu_2$ in the $\tau^+\tau^-$ CM frame:
\begin{equation}
s^\mu_1 = \left( \frac{\vec{k}_1 \cdot \vec{s}_1}{m}, \vec{s}_1 + \frac{\vec{k}_1 (\vec{k}_1 \cdot \vec{s}_1)}{m(m + k_1^0)} \right), \ \ \
s^\mu_2 = \left( \frac{\vec{k}_2 \cdot \vec{s}_2}{m}, \vec{s}_2 + \frac{\vec{k}_2 (\vec{k}_2 \cdot \vec{s}_2)}{m(m + k_2^0)} \right).
\end{equation}
Here  $m$ is the mass of $\tau$ lepton. $\vec{s}_1 = \frac{1}{2}(\vec{\sigma}\otimes\mathbb{1}$) and $\vec{s}_2 = \frac{1}{2}(\mathbb{1}\otimes\vec{\sigma}$) with $\mathbb{1}$ and  $\vec{\sigma} = (\sigma_1, \sigma_2, \sigma_3)$ being $2\times 2$ identity matrix and the Pauli matrices, respectively.  $k_1^0$ ($k_2^0$) and $\vec{k}_1$ ($\vec{k}_2$) are the energy and three-momentum components of the $\tau^-$ ($\tau^+$).
The leading order (LO) squared amplitudes for this subprocess have the following four terms:
\begin{equation}\label{eq:LO}
\begin{aligned}
\left | M_{\mathrm{LO}} \right |^2= \frac{64\,\alpha^2 \pi^2}{\bigl(1-\beta^2 y^2\bigr)^2\,m^{2}}
 & \Big[
A_0+B_0\,(p_1 \cdot s_2)(p_2 \cdot s_1)
+ C_0\,(p_1 \cdot s_1)(p_2 \cdot s_2)
+ D_0\,(s_1 \cdot s_2)\Big] ,
\end{aligned}
\end{equation}
with
\begin{equation}
\begin{aligned}
&A_0 = m^2\left(1-\beta^2\left[2(y^2-1)+\beta^2(2-2y^2+y^4)\right]\right),\\
&B_0 = \beta^2(\beta^2-1)(y^2-1)(-1+\beta y),\\
&C_0 = -\beta^2(\beta^2-1)(y^2-1)(1+\beta y),\\
&D_0 = m^2\Bigl(1-2\beta^2+\beta^4(2-2y^2+y^4)\Bigr),
\end{aligned}
\end{equation}
where $\beta=\sqrt{1-4m^2/s_{\gamma\gamma}}$ denotes the velocity of the $\tau$, and $s_{\gamma \gamma}=(p_1+p_2)^2$ is the squared CM energy of two photons. The variable $ y = \mathrm{cos}\theta $ with $\theta$ being the scattering angle between $p_1$ and $k_1$ in this frame.

The NLO EW results are rather lengthy and therefor not shown here. The unpolarized tau pair production in Pb--Pb UPC at NLO EW precision is already investigated in our previous work~\cite{Jiang:2024dhf}. In this paper, we focus on studying the $\tau^+\tau^-$ spin effects related to the process in Eq.~\eqref{eq:ggtautau} at NLO EW precision.

The production density matrix for $\gamma \gamma \to \tau^+ \tau^-$ process is defined by
\begin{eqnarray}\label{Rggtautau}
 R_{\alpha_1\alpha_2,\ \beta_1\beta_2}
 = \overline{\sum_{\lambda_1,\lambda_2}} &
 \langle \tau^-(k_1,\alpha_2),\, \tau^+(k_2,\beta_2) \vert
 \mathcal{T} \vert \gamma(p_1,\lambda_1),\, \gamma(p_2,\lambda_2) \rangle^*
 \nonumber \\[2mm]
 &\times
 \langle \tau^-(k_1,\alpha_1),\, \tau^+(k_2,\beta_1) \vert
 \mathcal{T} \vert \gamma(p_1,\lambda_1),\, \gamma(p_2,\lambda_2) \rangle \, ,
\end{eqnarray}
where the bar denotes averaging over the initial photon polarizations, and $\mathcal{T}$ is the transition operator.
The indices $\alpha$ and $\beta$ are spin labels referring to $\tau^-$ and $\tau^+$, respectively.
The decomposition of the matrix $R$ in the spin spaces of $\tau^-$ and $\tau^+$ is given by:
\begin{equation}\label{Rdec_tautau}
 R 
 = 
 \tilde  A\,\mathbb{1}\!\otimes\!\mathbb{1}
 + \tilde B_i^{\,+}\,\sigma^i\!\otimes\!\mathbb{1}
 + \tilde B_i^{\,-}\,\mathbb{1}\!\otimes\!\sigma^i
 + \tilde C_{ij}\,\sigma^i\!\otimes\!\sigma^j.
\end{equation}
The first factor in each tensor products refers to the spin space of $\tau^-$, and the second factor corresponds to that of $\tau^+$.
The scalar coefficient $\tilde A$ represents the unpolarized production rate, while the vector $\tilde B^{+}_i$ and $\tilde B^{-}_i$ correspond to the single spin polarization of $\tau^-$ and $\tau^+$, respectively. 
At LO in quantum field theory, the parity symmetry of the electromagnetic interaction ensures that the $\tilde B^{+}_i$ and $\tilde B^{-}_i$ vanish. 
Although weak corrections introduce parity-violating contributions, their numerical impact is negligible in comparison to $\tau$-pair spin correlations coefficients and is neglected in this paper.
The tensor $\tilde C_{ij}$ captures the spin correlations between the two $\tau$ leptons, and under the aforementioned approximation, the production density matrix is fully characterized by the coefficients $\tilde A$ and $\tilde C_{ij}$.

The tensor $\tilde C_{ij}$ can be further decomposed by introducing an orthonormal basis, which is defined as follows.
In what follows, $\hat{\bm{k}}$ denotes the flight direction of the $\tau^-$ in the $\tau^+\tau^-$ zero-momentum frame (ZMF), and the unit vector $\hat{\bm{p}} = \hat{\bm{p}}_1$ specifies the direction of one of the incoming photons in this frame.
The set of vectors introduced below then forms a right-handed orthonormal basis.~\cite{Bernreuther_2015}:
\begin{align}\label{orthoset_tautau}
 \bigl\{\,\hat{\bm{r}},\, \hat{\bm{k}},\, \hat{\bm{n}}\,\bigr\}:
 \qquad
 \hat{\bm{r}} = \frac{1}{r}\bigl(\hat{\bm{p}} - y\,\hat{\bm{k}}\bigr), \qquad
 \hat{\bm{n}} = \frac{1}{r}\bigl(\hat{\bm{p}} \times \hat{\bm{k}}\bigr), \qquad
 y = \hat{\bm{k}}\!\cdot\!\hat{\bm{p}}, \qquad
 r = \sqrt{1-y^2} \, .
\end{align}
We expand the tensor $\tilde C_{ij}$ with respect\footnote{The expansion of $\tilde C_{ij}$ in a chosen basis is not unique because $\delta_{ij} = \hat{n}_i\hat{n}_j + \hat{r}_i\hat{r}_j + \hat{k}_i\hat{k}_j$.} to the orthogonal basis in Eq.~\eqref{orthoset_tautau}, utilizing rotational invariance:
\begin{align}
 {\tilde C}_{ij} &=
   c_{rr}\,\hat{r}_i\hat{r}_j
 + c_{kk}\,\hat{k}_i\hat{k}_j
 + c_{nn}\,\hat{n}_i\hat{n}_j \notag \\[1mm]
 &\quad
 + c_{rk}\bigl(\hat{r}_i\hat{k}_j + \hat{k}_i\hat{r}_j\bigr)
 + c_{kn}\bigl(\hat{k}_i\hat{n}_j + \hat{n}_i\hat{k}_j\bigr)
 + c_{rn}\bigl(\hat{r}_i\hat{n}_j + \hat{n}_i\hat{r}_j\bigr) \notag \\[1mm]
 &\quad
 + \epsilon_{ijl}\bigl(
     c_r\,\hat{r}_l
   + c_k\,\hat{k}_l
   + c_n\,\hat{n}_l
   \bigr) \, . \label{eq:C_tautau}
\end{align}
The coefficients $c_{vv'}$ ($v,v' = r,k,n$)   are functions of $s_{\gamma \gamma}$ and $y$.
The terms in the antisymmetric part of Eq.~\eqref{eq:C_tautau} may also be written as
\begin{equation}\label{antiC_tautau}
 c_r\,\epsilon_{ijl}\hat{r}_l = c_r(\hat{k}_i\hat{n}_j - \hat{n}_i\hat{k}_j), \qquad
 c_k\,\epsilon_{ijl}\hat{k}_l = c_k(\hat{n}_i\hat{r}_j - \hat{r}_i\hat{n}_j), \qquad
 c_n\,\epsilon_{ijl}\hat{n}_l = c_n(\hat{r}_i\hat{k}_j - \hat{k}_i\hat{r}_j) \, .
\end{equation}

The production density matrix $R$ fully describes the $\tau^+\tau^-$ spin state.
 Connecting it to experimentally measurable quantities requires defining the spin correlation observables~\cite{Bernreuther:2004jv}
\begin{equation}
C_{ab}=4 \langle (\mathbf{\hat{a}}\!\cdot\!\vec{s}_1)\, (\mathbf{\hat{b}}\!\cdot\!\vec{s}_2) \rangle  = \frac{\int \phi  a_i \tilde{C}_{ij} b_j \ d \Pi _\mathrm{LIPS}}{\int d \sigma}. 
\label{eq:cab}
\end{equation}
Here, the flux factor $\phi=1/(2s_{\gamma \gamma})$, $a_i$ and $b_j$ are the $i$-th and $j$-th components of $\mathbf{\hat{a}}$ and $\mathbf{\hat{b}}$, respectively,
where the $\mathbf{\hat{a}}$ ($\mathbf{\hat{b}}$) are the chosen spin quantization axes for $\tau^-$ ($\tau^+$). 
$\Pi _\mathrm{LIPS}$ is the corresponding lorentz invariant phase space.
In this paper, they are those axes defined in Eq.~\eqref{orthoset_tautau}.
Eq.~\eqref{eq:cab} is defined with respect to the matrix elements for the production of $\tau$-pair, which is related to the more familiar double spin asymmetries \cite{PhysRevLett.87.242002,Bernreuther:2004jv}:
\begin{equation}
C_{ab} = \frac{\sigma(\uparrow \uparrow) + \sigma(\downarrow \downarrow ) - \sigma(\uparrow \downarrow ) - \sigma(\downarrow  \uparrow)}{\sigma(\uparrow \uparrow) + \sigma(\downarrow \downarrow ) + \sigma(\uparrow \downarrow  ) + \sigma(\downarrow  \uparrow)},
\label{eq:spin_corr}
\end{equation}
where the \( \uparrow \) and \( \downarrow  \) refer to the spin orientations of the \( \tau^+ \) and \( \tau^- \) along their respective spin quantization axis, and 
the first (second) arrow refers to the spin state of the \( \tau^+ \) (\( \tau^- \)).

\section{ Numerical results and analysis}\label{sec4}
In this section, we adopt $\alpha(0)$ scheme and  present numerical results for the NLO EW corrections to spin correlation observables in the process $\gamma\gamma \to \tau^+ \tau^-$ and to the total cross section Eq.\eqref{eq:sigma_total}, considering both Pb-Pb UPC and LC environments and then discuss the quantum entanglement of this system.
 
The input parameters adopted in our calculations are specified as follows~\cite{websitpdg}:
\begin{equation}
\begin{aligned}
\label{9}
&  ~~~~~~ m_W=80.369\ \mathrm{GeV},   ~~~~~~ m_Z=91.188 \ \mathrm{GeV}, ~~~~~~ m_H = 125.20\ \mathrm{GeV}, \\
&  ~~~~~~ m_t=172.4 \ \mathrm{GeV},   ~~~~~~ m_b=4.183 \ \mathrm{GeV}, ~~~~~~ m_c=1.273\ \mathrm{GeV},   \\
&  ~~~~~~ m_e=0.511 \ \mathrm{MeV},  ~~~~~~ m_\mu=0.106 \ \mathrm{GeV}, ~~~~~~  m=1.777 \ \mathrm{GeV}, ~~~~~~ \alpha^{-1}(0) = 137.036.
\end{aligned}
\end{equation}

\subsection{Total cross section and spin observables}\label{sec4.1}
We first discuss the inclusive cross section of the process and the spin observables. In Table~\ref{ac_UPC}, we show the cross section at various nucleon-nucleon CM energies   $\sqrt{s_{NN}}$, as available for Pb-Pb UPC at LHC~\cite{Bruce:2018yzs,dEnterria:2022sut}. The $\sigma_{\mathrm{LO}}$, $\delta\sigma_{\mathrm{NLO}}$ and $\sigma_{\mathrm{NLO}}$ denote the LO, pure NLO correction, and total NLO cross section, respectively and the factor $\delta=\frac{\sigma_{\mathrm{NLO}}}{\sigma_{\mathrm{LO}}}-1$ in the last column denotes the relative correction. These numerical results are given in the $\alpha(0)$  scheme and for a comparison, results in the $G_{\mu}$ scheme can be found in our previous work~\cite{Jiang:2024dhf}. 
Both results  are consistent with ref.~\cite{Dittmaier:2025ikh}, with minor discrepancies arising from difference in parameterization of photon flux and survival probability for the heavy ions.
For LC, we summarized the cross sections for $e^+e^-$ and $\mu^+\mu^-$ initial states  in Table~\ref{ac_ee} and Table~\ref{acMuon colliders} at various benchmark CM energies , respectively, following the same notation as Table~\ref{ac_UPC}.

\begin{table}[ht]
	\centering
	\begin{tabular}{ccccc}
		\hline
		$\sqrt{s_{NN}}$ [TeV] & $\sigma_{\mathrm{LO}}$ [mb] & $\delta\sigma_{\mathrm{NLO}} $ [mb] & $\sigma_{\mathrm{NLO}} $ [mb] & $\delta[\%]$ \\ \hline
		5.02 & 0.940 & $8.55\times 10^{-3}$ & 0.949 & 0.909 \\ 
		5.36 & 1.003 & $9.44\times 10^{-3}$ & 1.012 & 0.941 \\ 
		5.52 & 1.030 & $9.93\times 10^{-3}$ & 1.040 & 0.964 \\ \hline
	\end{tabular}
	\caption{Cross sections for $\gamma\gamma \to \tau^+\tau^-$ in Pb-Pb UPC at different nucleon-nucleon CM energies $\sqrt{s_{NN}}$. }
	\label{ac_UPC}
\end{table}

\begin{table}[ht]
	\centering
	\begin{tabular}{ccccc}
		\hline
		$\sqrt{s}$ [GeV] & $\sigma_{\mathrm{LO}}$ [nb] & $\delta\sigma_{\mathrm{NLO}}$ [nb] & $\sigma_{\mathrm{NLO}}$ [nb] & $\delta$ [\%] \\
		\hline
		250 & 0.5090 & $4.182\times 10^{-3}$ & 0.513 & 0.821 \\
		380  & 0.5350 & $4.178\times 10^{-3}$ & 0.539 & 0.781 \\
		1500 & 1.028  & $5.613\times 10^{-3}$ & 1.034 & 0.546 \\
		3000 & 1.340  & $3.499\times 10^{-3}$ & 1.344 & 0.261 \\
		\hline
	\end{tabular}
	\caption{Cross sections for $\gamma\gamma \to \tau^+\tau^-$ at $e^+e^-$ colliders at different $e^+e^-$  CM energies $\sqrt{s}$.} 
	\label{ac_ee}
\end{table}

\begin{table}[ht]
	\centering
	\begin{tabular}{ccccc}
		\hline
		$\sqrt{s}$ [TeV] & $\sigma_{\mathrm{LO}}$ [nb] & $\delta\sigma_{\mathrm{NLO}}$ [nb] & $\sigma_{\mathrm{NLO}}$ [nb] & $\delta$ [\%] \\
		\hline
		3  & 0.4990 & $1.807\times 10^{-3}$  & 0.501 & 0.362  \\
		10 & 0.8290 & $-3.656\times 10^{-3}$ & 0.826 & -0.441 \\
		\hline
	\end{tabular}
	\caption{Cross sections for $\gamma\gamma \to \tau^+\tau^-$ at $\mu^+\mu^-$ collider at different $\mu^+\mu^-$  CM energies $\sqrt{s}$.}
	\label{acMuon colliders}
\end{table}

For the UPC setup, as shown in Table~\ref{ac_UPC}, the total cross section at NLO EW is of the order of about 1 mb, with EW correction contributing approximately nine permille to LO at $\sqrt{s_{NN}}$ = 5.02 TeV, 5.36 TeV and 5.52 TeV.  
The relative correction factor $\delta$ in the last column increases with $\sqrt{s_{NN}}$, which means the EW effects become more significant at higher collision energies. 
On the contrary, for $e^+ e^-$ colliders, as presented in Table~\ref{ac_ee}, the relative contribution of NLO EW corrections decreases with CM energy. 
This trend shows that NLO EW corrections are more relevant at lower energies, while its impact diminishes in the high-energy regime considered. 
The total cross sections at NLO increase from 0.513 nb at $\sqrt{s}=250$ GeV to 1.344 nb at $\sqrt{s}=3$ TeV with the NLO EW correction also at about the permille level.
For Muon colliders, as shown in Table~\ref{acMuon colliders}, the total cross section is 0.501 nb at $\sqrt{s}$ = 3 TeV, corresponding to a 3.62 permille relative NLO contribution. 
The cross section difference from Table~\ref{ac_ee} at $\sqrt{s}=3$ TeV originates from the mass dependence of the PDF, resulting in a larger and more observable cross section for $e^{+}e^{-}$ colliders than for Muon colliders.
At $\sqrt{s}=10$ TeV, both the pure NLO correction and $\delta$ factor becomes negative, where the negative weak corrections dominate. 

We turn to discuss spin observables, evaluated with respect to defined set of reference axes.
The inclusive LO and NLO results for the spin correlation coefficients $C_{ab}$ are presented in Table~\ref{table_cab2} for UPC, while in Table~\ref{table_cab-ee} and Table~\ref{table_cab-Muon colliders} for $e^+e^-$ colliders and Muon collider, respectively. 
In these tables, we show the P-even and CP-even correlation coefficients such as $C_{rr}$, $C_{nn}$, $C_{kk}$ and $C_{pp}$. 
The P-odd or CP-odd spin correlation coefficients, which originate from parity-violating weak contributions, are not included as their contributions are numerially small and can be neglected safely. 

One can see from these tables, within the SM, the spin correlation coefficients $C_{ab}$ exhibit a characteristic pattern that can be qualitatively understood from angular momentum conservation and the dominance of specific partial waves in different energy regimes.
The behavior of $C_{rr}$ and $C_{kk}$ is particularly instructive. 
Near the $\tau^+\tau^-$ invariant mass threshold, the $\tau^+\tau^-$ system produced via photon fusion tends to be $^1S_0$ state, a configuration which will be shown in the following subsection to yield antiparallel spins along any common quantization axis.
When projected onto the individual $\hat{r}$-axes or $\hat{k}$-axes (the helicity axes), this antiparallel alignment manifests preferentially as $\uparrow\downarrow$ or $\downarrow\uparrow$ configurations, leading to negative values for both $C_{rr}$ and $C_{kk}$ in the threshold region.
As the CM energy increases, the $\tau$ leptons become relativistic, and helicity conservation becomes the dominant constraint, driving a transition toward a $^3S_1$-like state. 
Consequently, the spin projections along the $\hat{r}$- and helicity axes shift towards $\uparrow\uparrow$ or $\downarrow\downarrow$ configurations. 
This leads to an increase in the values of $C_{rr}$ and $C_{kk}$ at higher energies, although the increase differs between collider environments due to variations in the photon spectrum and polarization. 
The rather large value of $C_{nn}$ in the SM can be understood from total angular momentum conservation.
The consistently small magnitude of $C_{pp}$ across different energies and colliders suggests that spin correlations along the photon beam direction are generally suppressed.

The NLO EW corrections introduce only tiny modifications, primarily a slight rescaling of the correlation amplitudes without altering the dominant signs or energy-dependence trends. This provides a theoretically consistent and stable baseline for high-precision SM calculations.

\begin{table}[]
\begin{tabular}{|c|c|c|c|c|c|}
\hline
$\sqrt{s_{NN}}$ {[}TeV{]} & $C_{ab}$ & $C_{pp}$ & $C_{nn}$ & $C_{rr}$ & $C_{kk}$ \\ \hline
\multirow{2}{*}{5.02} & LO & -0.0213 & -0.3281 & 0.0199 & -0.2998 \\
\cline{2-6}
& NLO & -0.0214 & -0.3230 & 0.0195 & -0.2956 \\
\hline
\multirow{2}{*}{5.36} & LO & -0.0192 & -0.3273 & 0.0214 & -0.2968 \\ \cline{2-6}
& NLO & -0.0193 & -0.3223 & 0.0210 & -0.2926 \\
\hline
\multirow{2}{*}{5.52} & LO & -0.0183 & -0.3270 & 0.0221 & -0.2955\\ \cline{2-6}
& NLO & -0.0184 & -0.3220 & 0.0216 & -0.2913 \\
\hline
\end{tabular}
\caption{The spin correlation coefficients $C_{ij}$ in Pb-Pb UPC at LO and NLO EW at different nucleon-nucleon CM energies $\sqrt{s_{NN}}$.}\label{table_cab2}
\end{table}

\begin{table}[]
\begin{tabular}{|c|c|c|c|c|c|}
\hline
$\sqrt{s}$ {[}GeV{]} & $C_{ab}$ & $C_{pp}$& $C_{nn}$ & $C_{rr}$ & $C_{kk}$ \\ \hline
\multirow{2}{*}{250} & LO & 0.0335 & -0.3090 & 0.0580 & -0.2183 \\ \cline{2-6} 
                    & NLO & 0.0334 & -0.3070 & 0.0579 & -0.2167 \\ \hline
\multirow{2}{*}{380} & LO & 0.0514 & -0.3030 & 0.0695 & -0.1909 \\ \cline{2-6} 
                    & NLO & 0.0513 & -0.3012 & 0.0694 & -0.1893 \\ \hline
\multirow{2}{*}{1500} & LO & 0.0705 & -0.2973 & 0.0796 & -0.1604 \\ \cline{2-6} 
                    & NLO & 0.0703 & -0.2956 & 0.0795 & -0.1590 \\ \hline
\multirow{2}{*}{3000} & LO & 0.0763 & -0.2956 & 0.0825 & -0.1512 \\ \cline{2-6} 
                    & NLO & 0.0756 & -0.2938 & 0.0824 & -0.1506 \\ \hline
\end{tabular}
\caption{The spin correlation coefficients $C_{ij}$ for $e^+e^-$ colliders at LO and NLO EW at different $e^+e^-$ CM energies $\sqrt{s}$.}\label{table_cab-ee}
\end{table}

\begin{table}[]
\begin{tabular}{|c|c|c|c|c|c|}
\hline
$\sqrt{s}$ {[}TeV{]} & $C_{ab}$ & $C_{pp}$ & $C_{nn}$ & $C_{rr}$ & $C_{kk}$ \\ \hline
\multirow{2}{*}{3} & LO & 0.0743 & -0.2905 & 0.0799 & -0.1515 \\ \cline{2-6} 
& NLO & 0.0737 & -0.2944 & 0.0814 & -0.1536 \\ \hline
\multirow{2}{*}{10} & LO & 0.0821 & -0.2938 & 0.0853 & -0.1419 \\ \cline{2-6} 
& NLO & 0.0799 & -0.2921 & 0.0852 & -0.1533\\ \hline
\end{tabular}
\caption{The spin correlation coefficients $C_{ij}$ for Muon colliders at LO and NLO EW at different $\mu^+\mu^-$ CM energies $\sqrt{s}$.}\label{table_cab-Muon colliders}
\end{table}

\subsection{Quantum Entanglement} 
It is natural to investigate whether these correlations arise solely from classical spin alignment or contain genuine quantum entanglement features, with the establishment of the spin correlation coefficients characterizing the joint spin configurations of the $\tau^+\tau^-$ system.
The extent of quantum entanglement can be quantified by the concurrence~\cite{Wootters:1997id}.

For the process we considered, in the nonrelativistic limit, the concurrence can be written as:
\begin{equation}
\mathcal{C}(\rho) = -\frac{1}{2}(1 + C_{11} + C_{22} + C_{33})
                 = -\frac{1}{2}(1 + 3D),
\end{equation}
where the scalar quantity
\begin{equation}
D = \frac{C_{11} + C_{22} + C_{33}}{3} = \frac{\mathrm{tr}[C]}{3}
\label{eq:D_def}
\end{equation}
serves as a compact, rotation invariant criterion for entanglement, and $C_{11}, C_{22}, C_{33}$ represent the spin correlation coefficients corresponding to any set of orthogonal vectors.
The sufficient condition for an entangled $\tau^+\tau^-$ state is $-1 \leqslant D < -\tfrac{1}{3}$, with $D=-1$ corresponding to maximal entanglement.
Thus, quantum entanglement can be probed through a single spin correlation observable, as has been demonstrated in $t\bar t$ analyses at the LHC~\cite{ATLAS:2023fsd,CMS:2024pts}.
$D$ is invariant under rotation and can therefore be evaluated in any orthonormal basis, such as
$\{\hat{\bm{n}}, \hat{\bm{r}}, \hat{\bm{k}}\}$ defined in Eq.~\eqref{orthoset_tautau}.
Our simulation across the inclusive phase space shows that the average value of quantum entanglement criterion $D$, is always greater than $-1/3$, thereby failing to satisfy the entanglement condition. 
In realistic analyses, NLO EW corrections to $\tau$ decays do influence its distributions, and potentially affect the spin-correlation observables such as $D$. However, this influence is minor, amounting to only a few percent as shown in  ref.~\cite{Dittmaier:2025ikh}.

\begin{figure}[thbp]
	\centering
	\includegraphics[width=0.6\textwidth]{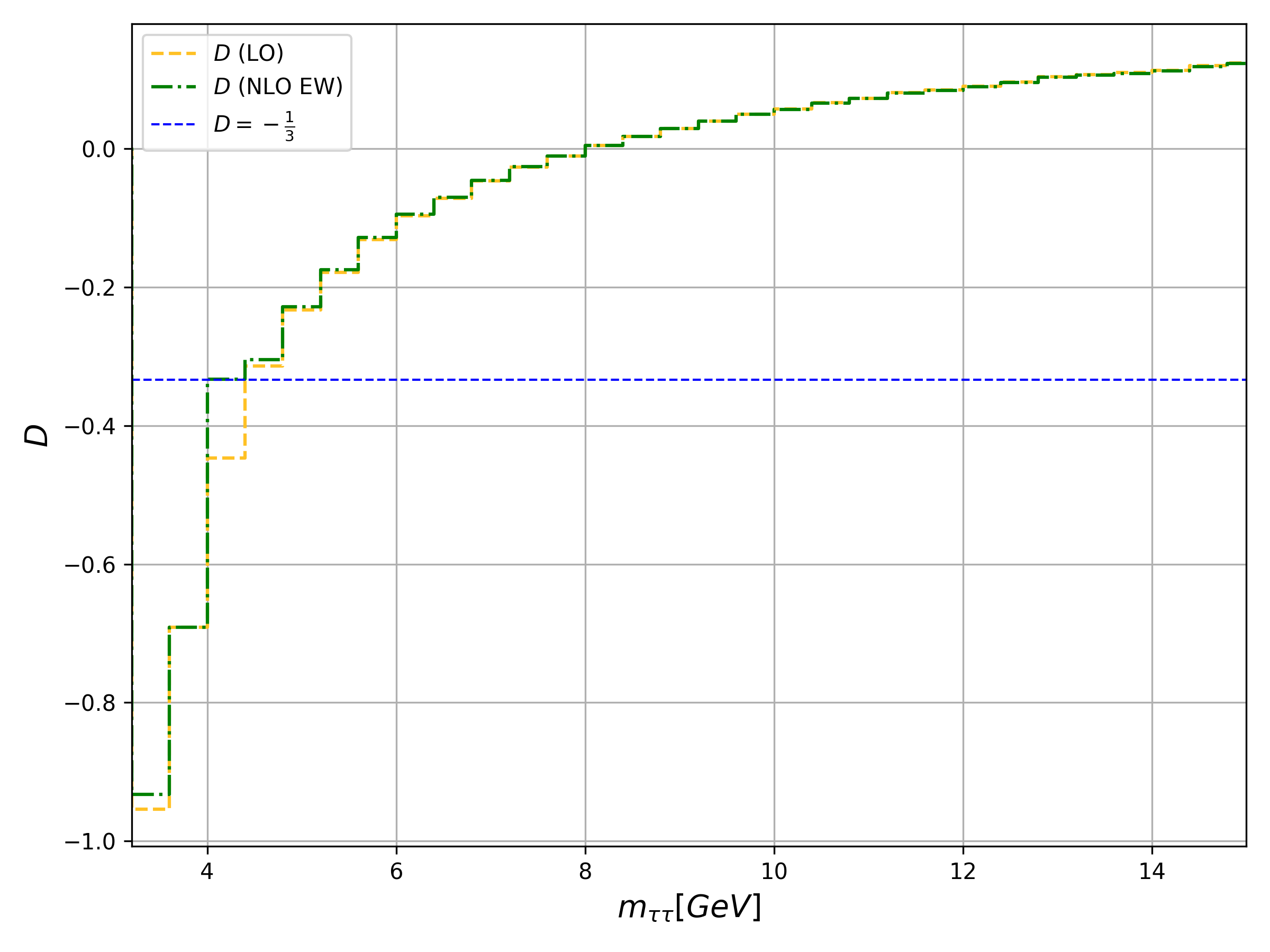}
\caption{
The differential entanglement observable $D$ is shown as a function of the $\tau$-pair invariant mass $m_{\tau\tau}$ for $\sqrt{s_{NN}} = 5.02$ TeV in UPC.
The representative mass interval displayed is  $m_{\tau\tau}\in[2 m, 15]$~GeV.
The yellow dashed and  green dash-dotted curves represent the LO and NLO EW results, respectively.
The horizontal blue dashed line indicates the entanglement threshold at $D=-1/3$.
Negative values near  $\tau^+\tau^-$ invariant mass threshold reflect the dominance of spin singlet configurations, while the transition toward positive $D$ at higher masses signals the onset of spin-triplet states and reduced entanglement.
}
	\label{fig:D}
\end{figure}

To further explore quantum entanglement event-by-event,
the observable $D$ can be studied as a function of the invariant mass of the produced $\tau$-pair, defined as $m_{\tau\tau}$.
Fig.~\ref{fig:D} plots the value of $D$ for each bin as a function of the $m_{\tau\tau}$ in the interval $[2 m, 15]$ GeV.
The  yellow dashed line shows the LO result, and the green dash-dotted line includes NLO EW corrections.
The horizontal blue dashed line marks the entanglement threshold $D =-1/3$.
Near $\tau^+\tau^-$ invariant mass threshold, both curves exhibit large negative values of $D$,  well below the entanglement criterion, which originates from the dominance of low-$\beta$ configurations. 
In this regime, the $\tau^+\tau^-$ system produced via photon fusion is predominantly in a spin singlet state $^1S_0$, resulting in antiparallel spins along any common quantization axis and representing a maximally quantum entangled state.
The NLO EW corrections induce modest upward shift while preserving the overall shape, slightly reducing the degree of entanglement near $\tau^+\tau^-$ invariant mass threshold.

\section{Summary}\label{sec5}

In this paper, we present a comprehensive analysis of the process $\gamma\gamma \to \tau^+\tau^-$ including $\tau^+ \tau^-$ spin information at NLO EW accuracy for both in Pb-Pb UPC and at  LC.
We provide the predictions for  the corresponding cross section and the spin correlation coefficients $C_{ab}$.
For Pb--Pb UPC, the NLO EW corrected cross sections are found to be of the order of $1$~mb with positive NLO EW corrections at the level of a few permille in the energy range we considered. 
For LC, $e^+e^-$ collider has superior observability over  Muon collider at equal CM energy.
For the spin correlation coefficients, $C_{rr}$ and $C_{kk}$ tend to take negative values close to $\tau^+\tau^-$ invariant mass threshold where antiparallel spin configurations dominate, and
 $C_{nn}$ remains large in magnitude due to total angular momentum conservation, whereas $C_{pp}$ stays consistently small. 
The NLO EW corrections produce small numerical shifts and do not alter these qualitative features.  
Using the available  spin coefficients $C_{nn}$, $C_{rr}$ and $C_{kk}$, we investigate quantum entanglement in the $\tau^+\tau^-$ system by evaluating the rotation invariant entanglement criterion $D$.
The inclusive value of $D$ remains above the entanglement threshold $D=-1/3$, indicating no net entanglement after full phase space integration. 
However, a differential study as a function of the invariant mass $m_{\tau\tau}$ reveals that  $D<-1/3$ near the threshold region where the total spin configuration of $\tau^+\tau^-$ tends to be $^1S_0$ state, so that a genuinely entangled quantum state can be confirmed.
The study in this paper can provide a helpful analysis template to  study the   $\tau$-pair production induced by $\gamma\gamma$ collision at high energy colliders.

\section*{Acknowledgements}

The authors thank to the members of the Institute of theoretical physics of Shandong University for their helpful communications. This work is supported in part by National Natural Science Foundation of China under the Grants    No.12321005, No. 12235008 and No.12405121.
\newpage

\bibliography{ref}

@article{Bernreuther_2015,
   title={A set of top quark spin correlation and polarization observables for the LHC: Standard Model predictions and new physics contributions},
   volume={2015},
   ISSN={1029-8479},
   url={http://dx.doi.org/10.1007/JHEP12(2015)026},
   DOI={10.1007/jhep12(2015)026},
   number={12},
   journal={Journal of High Energy Physics},
   publisher={Springer Science and Business Media LLC},
   author={Bernreuther, Werner and Heisler, Dennis and Si, Zong-Guo},
   year={2015},
   month=dec, pages={1–36} }

@article{vonWeizsacker:1934nji,
    author = "von Weizsacker, C. F.",
    title = "{Radiation emitted in collisions of very fast electrons}",
    doi = "10.1007/BF01333110",
    journal = "Z. Phys.",
    volume = "88",
    pages = "612--625",
    year = "1934"
}

@article{ParticleDataGroup:2024cfk,
    author = "Navas, S. and others",
    collaboration = "Particle Data Group",
    title = "{Review of particle physics}",
    doi = "10.1103/PhysRevD.110.030001",
    journal = "Phys. Rev. D",
    volume = "110",
    number = "3",
    pages = "030001",
    year = "2024"
}

@article{PhysRevLett.87.242002,
  title = {Top-Quark Spin Correlations at Hadron Colliders: Predictions at Next-to-Leading Order QCD},
  author = {Bernreuther, W. and Brandenburg, A. and Si, Z. G. and Uwer, P.},
  journal = {Phys. Rev. Lett.},
  volume = {87},
  issue = {24},
  pages = {242002},
  numpages = {4},
  year = {2001},
  month = {Nov},
  publisher = {American Physical Society},
  doi = {10.1103/PhysRevLett.87.242002},
  url = {https://link.aps.org/doi/10.1103/PhysRevLett.87.242002}
}

@article{ATLAS:2022ryk,
    author = "Aad, Georges and others",
    collaboration = "ATLAS",
    title = "{Observation of the {\ensuremath{\gamma}}{\ensuremath{\gamma}}{\textrightarrow}{\ensuremath{\tau}}{\ensuremath{\tau}} Process in Pb+Pb Collisions and Constraints on the {\ensuremath{\tau}}-Lepton Anomalous Magnetic Moment with the ATLAS Detector}",
    eprint = "2204.13478",
    archivePrefix = "arXiv",
    primaryClass = "hep-ex",
    reportNumber = "CERN-EP-2022-079",
    doi = "10.1103/PhysRevLett.131.151802",
    journal = "Phys. Rev. Lett.",
    volume = "131",
    number = "15",
    pages = "151802",
    year = "2023"
}

@article{CMS:2024qjo,
    author = "Hayrapetyan, Aram and others",
    collaboration = "CMS",
    title = "{Observation of $\gamma\gamma\to\tau\tau$ in proton-proton collisions and limits on the anomalous electromagnetic moments of the $\tau$ lepton}",
    eprint = "2406.03975",
    archivePrefix = "arXiv",
    primaryClass = "hep-ex",
    reportNumber = "CMS-SMP-23-005, CERN-EP-2024-127",
    doi = "10.1088/1361-6633/ad6fcb",
    journal = "Rept. Prog. Phys.",
    volume = "87",
    number = "10",
    pages = "107801",
    year = "2024"
}

@article{ALEPH:2005qgp,
    author = "Schael, S. and others",
    collaboration = "ALEPH",
    title = "{Branching ratios and spectral functions of tau decays: Final ALEPH measurements and physics implications}",
    eprint = "hep-ex/0506072",
    archivePrefix = "arXiv",
    doi = "10.1016/j.physrep.2005.06.007",
    journal = "Phys. Rept.",
    volume = "421",
    pages = "191--284",
    year = "2005"
}

@article{Jiang:2024dhf,
    author = "Jiang, Jun and Lu, Peng-Cheng and Si, Zong-Guo and Zhang, Han and Zhang, Xin-Yi",
    title = "{NLO EW corrections to tau pair production via photon fusion in Pb-Pb ultraperipheral collisions}",
    eprint = "2410.21963",
    archivePrefix = "arXiv",
    primaryClass = "hep-ph",
    doi = "10.1103/PhysRevD.111.036023",
    journal = "Phys. Rev. D",
    volume = "111",
    number = "3",
    pages = "036023",
    year = "2025"
}

@article{Shao:2024dmk,
    author = "Shao, Hua-Sheng and d'Enterria, David",
    title = "{Dimuon and ditau production in photon-photon collisions at next-to-leading order in QED}",
    eprint = "2407.13610",
    archivePrefix = "arXiv",
    primaryClass = "hep-ph",
    doi = "10.1007/JHEP02(2025)023",
    journal = "JHEP",
    volume = "02",
    pages = "023",
    year = "2025"
}

@article{Dittmaier:2025ikh,
    author = "Dittmaier, Stefan and Engel, Tim and Ariza, Jose Luis Hernando and Pellen, Mathieu",
    title = "{Electroweak corrections to $\tau^+\tau^-$ production in ultraperipheral heavy-ion collisions at the LHC}",
    eprint = "2504.11391",
    archivePrefix = "arXiv",
    primaryClass = "hep-ph",
    reportNumber = "FR-PHENO-2025-004",
    month = "4",
    year = "2025"
}

@article{Dyndal:2023sts,
    author = "Dyndal, Mateusz",
    collaboration = "ALICE, ATLAS, CMS, LHCb, MoEDAL",
    title = "{Ultra-peripheral collisions (experiment)}",
    doi = "10.22323/1.422.0215",
    journal = "PoS",
    volume = "LHCP2022",
    pages = "215",
    year = "2023"
}

@article{Williams:1934ad,
    author = "Williams, E. J.",
    title = "{Nature of the high-energy particles of penetrating radiation and status of ionization and radiation formulae}",
    doi = "10.1103/PhysRev.45.729",
    journal = "Phys. Rev.",
    volume = "45",
    pages = "729--730",
    year = "1934"
}

@article{Frixione:1993yw,
    author = "Frixione, Stefano and Mangano, Michelangelo L. and Nason, Paolo and Ridolfi, Giovanni",
    title = "{Improving the Weizsacker-Williams approximation in electron - proton collisions}",
    eprint = "hep-ph/9310350",
    archivePrefix = "arXiv",
    reportNumber = "CERN-TH-7032-93, GEF-TH-18-93",
    doi = "10.1016/0370-2693(93)90823-Z",
    journal = "Phys. Lett. B",
    volume = "319",
    pages = "339--345",
    year = "1993"
}

@article{Shao:2022cly,
    author = "Shao, Hua-Sheng and d'Enterria, David",
    title = "{gamma-UPC: automated generation of exclusive photon-photon processes in ultraperipheral proton and nuclear collisions with varying form factors}",
    eprint = "2207.03012",
    archivePrefix = "arXiv",
    primaryClass = "hep-ph",
    doi = "10.1007/JHEP09(2022)248",
    journal = "JHEP",
    volume = "09",
    pages = "248",
    year = "2022"
}

@article{CMS:2022arf,
    author = "Tumasyan, Armen and others",
    collaboration = "CMS",
    title = "{Observation of $\tau$ lepton pair production in ultraperipheral lead-lead collisions at $\sqrt{s_\mathrm{NN}}$ = 5.02 TeV}",
    eprint = "2206.05192",
    archivePrefix = "arXiv",
    primaryClass = "nucl-ex",
    reportNumber = "CMS-HIN-21-009, CERN-EP-2022-098",
    doi = "10.1103/PhysRevLett.131.151803",
    journal = "Phys. Rev. Lett.",
    volume = "131",
    pages = "151803",
    year = "2023"
}

@book{Jackson:1998nia,
    author = "Jackson, John David",
    title = "{Classical Electrodynamics}",
    isbn = "978-0-471-30932-1",
    publisher = "Wiley",
    year = "1998"
}

@article{Baur:2001jj,
    author = "Baur, Gerhard and Hencken, Kai and Trautmann, Dirk and Sadovsky, Serguei and Kharlov, Yuri",
    title = "{Coherent gamma gamma and gamma-A interactions in very peripheral collisions at relativistic ion colliders}",
    eprint = "hep-ph/0112211",
    archivePrefix = "arXiv",
    doi = "10.1016/S0370-1573(01)00101-6",
    journal = "Phys. Rept.",
    volume = "364",
    pages = "359--450",
    year = "2002"
}

@article{Cahn:1990jk,
    author = "Cahn, Robert N. and Jackson, John David",
    title = "{Realistic equivalent photon yields in heavy ion collisions}",
    reportNumber = "LBL-28592-REV, LBL-28592",
    doi = "10.1103/PhysRevD.42.3690",
    journal = "Phys. Rev. D",
    volume = "42",
    pages = "3690--3695",
    year = "1990"
}

@article{Budnev:1975poe,
    author = "Budnev, V. M. and Ginzburg, I. F. and Meledin, G. V. and Serbo, V. G.",
    title = "{The Two photon particle production mechanism. Physical problems. Applications. Equivalent photon approximation}",
    doi = "10.1016/0370-1573(75)90009-5",
    journal = "Phys. Rept.",
    volume = "15",
    pages = "181--281",
    year = "1975"
}

@article{Klasen:2001cu,
    author = "Klasen, M. and Kniehl, Bernd A. and Mihaila, L. N. and Steinhauser, M.",
    title = "{Evidence for color octet mechanism from CERN LEP-2 $\gamma \gamma \to J/\psi$ + $X$ data}",
    eprint = "hep-ph/0112259",
    archivePrefix = "arXiv",
    reportNumber = "DESY-01-202",
    doi = "10.1103/PhysRevLett.89.032001",
    journal = "Phys. Rev. Lett.",
    volume = "89",
    pages = "032001",
    year = "2002"
}

@article{PhysRevLett.23.880,
  title = {Proposed Experiment to Test Local Hidden-Variable Theories},
  author = {Clauser, John F. and Horne, Michael A. and Shimony, Abner and Holt, Richard A.},
  journal = {Phys. Rev. Lett.},
  volume = {23},
  issue = {15},
  pages = {880--884},
  numpages = {0},
  year = {1969},
  month = {Oct},
  publisher = {American Physical Society},
  doi = {10.1103/PhysRevLett.23.880},
  url = {https://link.aps.org/doi/10.1103/PhysRevLett.23.880}
}

@misc{han2025entanglementbellnonlocalitytau,
      title={Entanglement and Bell Nonlocality in $\tau^+ \tau^-$ at the BEPC}, 
      author={Tao Han and Matthew Low and Youle Su},
      year={2025},
      eprint={2501.04801},
      archivePrefix={arXiv},
      primaryClass={hep-ph},
      url={https://arxiv.org/abs/2501.04801}, 
}

@article{CEPCStudyGroup:2018ghi,
    author = "Dong, Mingyi and others",
    editor = "Guimar{\~a}es da Costa, Jo{\~a}o Barreiro and others",
    collaboration = "CEPC Study Group",
    title = "{CEPC Conceptual Design Report: Volume 2 - Physics {\&} Detector}",
    eprint = "1811.10545",
    archivePrefix = "arXiv",
    primaryClass = "hep-ex",
    reportNumber = "IHEP-CEPC-DR-2018-02, IHEP-EP-2018-01, IHEP-TH-2018-01",
    month = "11",
    year = "2018"
}

@article{CLIC:2018fvx,
    author = "de Blas, J. and others",
    collaboration = "CLIC",
    title = "{The CLIC Potential for New Physics}",
    eprint = "1812.02093",
    archivePrefix = "arXiv",
    primaryClass = "hep-ph",
    reportNumber = "CERN-TH-2018-267, CERN-2018-009-M, FERMILAB-TM-2795",
    doi = "10.23731/CYRM-2018-003",
    journal = "CERN Yellow Rep. Monogr.",
    volume = "3",
    pages = "1--282",
    year = "2018"
}

@article{Delahaye:2019omf,
    author = "Delahaye, Jean Pierre and Diemoz, Marcella and Long, Ken and Mansouli{\'e}, Bruno and Pastrone, Nadia and Rivkin, Lenny and Schulte, Daniel and Skrinsky, Alexander and Wulzer, Andrea",
    title = "{Muon Colliders}",
    eprint = "1901.06150",
    archivePrefix = "arXiv",
    primaryClass = "physics.acc-ph",
    month = "1",
    year = "2019"
}

@article{MuonCollider:2022ded,
    author = "Bartosik, N. and others",
    collaboration = "Muon Collider",
    title = "{Simulated Detector Performance at the Muon Collider}",
    eprint = "2203.07964",
    archivePrefix = "arXiv",
    primaryClass = "hep-ex",
    reportNumber = "FERMILAB-FN-1185-AD-ND-PPD-TD",
    doi = "10.2172/1886011",
    month = "3",
    year = "2022"
}

@article{Wootters:1997id, 
	author = "Wootters, William K.", 
	title = "{Entanglement of formation of an arbitrary state of two qubits}", 
	eprint = "quant-ph/9709029", 
	archivePrefix = "arXiv", 
	doi = "10.1103/PhysRevLett.80.2245", 
	journal = "Phys. Rev. Lett.", 
	volume = "80", 
	pages = "2245--2248", 
	year = "1998" 
}

@article{Zhang:2025mmm,
    author = "Zhang, Yulei and Zhou, Bai-Hong and Liu, Qi-Bin and Li, Shu and Hsu, Shih-Chieh and Han, Tao and Low, Matthew and Wu, Tong Arthur",
    title = "{Entanglement and Bell Nonlocality in $\tau^+ \tau^-$ at the LHC using Machine Learning for Neutrino Reconstruction}",
    eprint = "2504.01496",
    archivePrefix = "arXiv",
    primaryClass = "hep-ph",
    month = "4",
    year = "2025"
}

@article{Wu:2024ovc,
    author = "Wu, Youpeng and Jiang, Ruobing and Ruzi, Alim and Ban, Yong and Yan, Xueqing and Li, Qiang",
    title = "{Testing Bell inequalities and probing quantum entanglement at CEPC}",
    eprint = "2410.17025",
    archivePrefix = "arXiv",
    primaryClass = "hep-ph",
    doi = "10.1103/PhysRevD.111.036008",
    journal = "Phys. Rev. D",
    volume = "111",
    number = "3",
    pages = "036008",
    year = "2025"
}

@article{ALEPH:2001uca, 
	author = "Heister, A. and others", 
	collaboration = "ALEPH", 
	title = "{Measurement of the tau polarization at LEP}", 
	eprint = "hep-ex/0104038", 
	archivePrefix = "arXiv", 
	reportNumber = "CERN-EP-2001-027",
	doi = "10.1007/s100520100689", 
	journal = "Eur. Phys. J. C", 
	volume = "20", 
	pages = "401--430", 
	year = "2001" 
}

@article{DELPHI:2003zcz, 
	author = "Abdallah, J. and others", 
	collaboration = "DELPHI", 
	title = "{A Precise measurement of the tau lifetime}", 
	eprint = "hep-ex/0410010", 
	archivePrefix = "arXiv", 
	reportNumber = "CERN-EP-2003-070", 
	doi = "10.1140/epjc/s2004-01953-7", 
	journal = "Eur. Phys. J. C", 
	volume = "36", 
	pages = "283--296", 
	year = "2004" 
}

@article{Belle:2006qqw,
	author = "Abe, Kazuo and others", 
	editor = "Sissakian, Alexey and Kozlov, Gennady and Kolganova, Elena", 
	collaboration = "Belle", 
	title = "{Measurement of the mass of the tau-lepton and an upper limit on the mass difference between tau+ and tau-}",
	eprint = "hep-ex/0608046", 
	archivePrefix = "arXiv", 
	reportNumber = "BELLE-CONF-0657", 
	doi = "10.1103/PhysRevLett.99.011801", 
	journal = "Phys. Rev. Lett.", 
	volume = "99", 
	pages = "011801", 
	year = "2007"
}

@article{Zhang:2018gol, 
	author = "Zhang, J. Y.", 
	collaboration = "BESIII", 
	title = "{$\tau$ lepton mass measurement at BESIII}", 
	eprint = "1812.10056", 
	archivePrefix = "arXiv", 
	primaryClass = "hep-ex", 
	doi = "10.21468/SciPostPhysProc.1.004", 
	journal = "SciPost Phys. Proc.", 
	volume = "1", 
	pages = "004", 
	year = "2019" 
}

@article{Cheng:2024rxi,
    author = "Cheng, Kun and Han, Tao and Low, Matthew",
    title = "{Quantum tomography at colliders: With or without decays}",
    eprint = "2410.08303",
    archivePrefix = "arXiv",
    primaryClass = "hep-ph",
    reportNumber = "PITT-PACC-2408",
    doi = "10.1016/j.physletb.2025.139675",
    journal = "Phys. Lett. B",
    volume = "868",
    pages = "139675",
    year = "2025"
}

@article{Ashby-Pickering:2022umy,
    author = "Ashby-Pickering, Rachel and Barr, Alan J. and Wierzchucka, Agnieszka",
    title = "{Quantum state tomography, entanglement detection and Bell violation prospects in weak decays of massive particles}",
    eprint = "2209.13990",
    archivePrefix = "arXiv",
    primaryClass = "quant-ph",
    doi = "10.1007/JHEP05(2023)020",
    journal = "JHEP",
    volume = "05",
    pages = "020",
    year = "2023"
}

@article{Han:2025ewp,
    author = "Han, Tao and Low, Matthew and Su, Youle",
    title = "{Entanglement and Bell Nonlocality in $τ^+ τ^-$ at the BEPC}",
    eprint = "2501.04801",
    archivePrefix = "arXiv",
    primaryClass = "hep-ph",
    reportNumber = "PITT-PACC-2412",
    month = "1",
    year = "2025"
}

@article{CMS:2024pts,
    author = "Hayrapetyan, Aram and others",
    collaboration = "CMS",
    title = "{Observation of quantum entanglement in top quark pair production in proton{\textendash}proton collisions at $\sqrt{s} = 13$ TeV}",
    eprint = "2406.03976",
    archivePrefix = "arXiv",
    primaryClass = "hep-ex",
    reportNumber = "CMS-TOP-23-001, CERN-EP-2024-137",
    doi = "10.1088/1361-6633/ad7e4d",
    journal = "Rept. Prog. Phys.",
    volume = "87",
    number = "11",
    pages = "117801",
    year = "2024"
}

@article{ATLAS:2023fsd,
    author = "Aad, Georges and others",
    collaboration = "ATLAS",
    title = "{Observation of quantum entanglement with top quarks at the ATLAS detector}",
    eprint = "2311.07288",
    archivePrefix = "arXiv",
    primaryClass = "hep-ex",
    reportNumber = "CERN-EP-2023-230",
    doi = "10.1038/s41586-024-07824-z",
    journal = "Nature",
    volume = "633",
    number = "8030",
    pages = "542--547",
    year = "2024"
}

@article{Fabbrichesi:2024wcd,
    author = "Fabbrichesi, M. and Marzola, L.",
    title = "{Quantum tomography with {\ensuremath{\tau}} leptons at the FCC-ee: Entanglement, Bell inequality violation, sin{\ensuremath{\theta}}W, and anomalous couplings}",
    eprint = "2405.09201",
    archivePrefix = "arXiv",
    primaryClass = "hep-ph",
    doi = "10.1103/PhysRevD.110.076004",
    journal = "Phys. Rev. D",
    volume = "110",
    number = "7",
    pages = "076004",
    year = "2024"
}

@article{Ehataht:2023zzt,
    author = {Ehat{\"a}ht, Karl and Fabbrichesi, Marco and Marzola, Luca and Veelken, Christian},
    title = "{Probing entanglement and testing Bell inequality violation with e+e-{\textrightarrow}{\ensuremath{\tau}}+{\ensuremath{\tau}}- at Belle II}",
    eprint = "2311.17555",
    archivePrefix = "arXiv",
    primaryClass = "hep-ph",
    doi = "10.1103/PhysRevD.109.032005",
    journal = "Phys. Rev. D",
    volume = "109",
    number = "3",
    pages = "032005",
    year = "2024"
}

@misc{websitpdg,
    howpublished = "\url{https://pdg.lbl.gov/2024/reviews/contents_sports.html}"
}

@article{ATLAS:2018ynr, 
    author = "Aaboud, Morad and others", 
    collaboration = "ATLAS", 
    title = "{Cross-section measurements of the Higgs boson decaying into a pair of $\tau$-leptons in proton-proton collisions at $\sqrt{s}=13$ TeV with the ATLAS detector}", 
    eprint = "1811.08856", 
    archivePrefix = "arXiv", 
    primaryClass = "hep-ex", 
    reportNumber = "CERN-EP-2018-232", 
    doi = "10.1103/PhysRevD.99.072001", 
    journal = "Phys. Rev. D", 
    volume = "99", 
    pages = "072001", 
    year = "2019" 
}

@article{ATLAS:2022akr, 
    author = "Aad, Georges and others", 
    collaboration = "ATLAS", 
    title = "{Measurement of the CP properties of Higgs boson interactions with $\tau $-leptons with the ATLAS detector}", 
    eprint = "2212.05833", 
    archivePrefix = "arXiv", 
    primaryClass = "hep-ex", 
    reportNumber = "CERN-EP-2022-244", 
    doi = "10.1140/epjc/s10052-023-11583-y", 
    journal = "Eur. Phys. J. C", 
    volume = "83", 
    number = "7", 
    pages = "563", 
    year = "2023" 
}

@article{Tsai:1971vv,
    author = "Tsai, Yung-Su",
    title = "{Decay Correlations of Heavy Leptons in e+ e- ---{\ensuremath{>}} Lepton+ Lepton-}",
    reportNumber = "SLAC-PUB-0932",
    doi = "10.1103/PhysRevD.13.771",
    journal = "Phys. Rev. D",
    volume = "4",
    pages = "2821",
    year = "1971",
    note = "[Erratum: Phys.Rev.D 13, 771 (1976)]"
}

@article{Kuhn:1990ad,
    author = "Kuhn, Johann H. and Santamaria, A.",
    title = "{Tau decays to pions}",
    reportNumber = "MPI-PAE/PTh-17/90",
    doi = "10.1007/BF01572024",
    journal = "Z. Phys. C",
    volume = "48",
    pages = "445--452",
    year = "1990"
}

@article{Bernabeu:2007rr,
    author = "Bernabeu, J. and Gonzalez-Sprinberg, G. A. and Papavassiliou, J. and Vidal, J.",
    title = "{Tau anomalous magnetic moment form-factor at super B/flavor factories}",
    eprint = "0707.2496",
    archivePrefix = "arXiv",
    primaryClass = "hep-ph",
    reportNumber = "FTUV-07-1607",
    doi = "10.1016/j.nuclphysb.2007.09.001",
    journal = "Nucl. Phys. B",
    volume = "790",
    pages = "160--174",
    year = "2008"
}

@article{Chen:2018cxt,
        author = "Chen, Xin and Wu, Yongcheng",
        title = "{Search for the Electric Dipole Moment and anomalous magnetic moment of the tau lepton at tau factories}",
        eprint = "1803.00501",
        archivePrefix = "arXiv",
        primaryClass = "hep-ph",
        doi = "10.1007/JHEP10(2019)089",
        journal = "JHEP",
        volume = "10",
        pages = "089",
        year = "2019"
}

@article{Bodrov:2024wrw,
    author = "Bodrov, Denis",
    title = "{Tau physics at Belle and Belle~II}",
    eprint = "2405.16512",
    archivePrefix = "arXiv",
    primaryClass = "hep-ex",
    doi = "10.1142/S0217751X24420065",
    journal = "Int. J. Mod. Phys. A",
    volume = "39",
    number = "26n27",
    pages = "2442006",
    year = "2024"
}

@article{Shao:2025bma,
    author = "Shao, Hua-Sheng and Simon, Lukas",
    title = "{Automated next-to-leading order QCD and electroweak predictions of photon-photon processes in ultraperipheral collisions}",
    eprint = "2504.10104",
    archivePrefix = "arXiv",
    primaryClass = "hep-ph",
    doi = "10.1007/JHEP07(2025)020",
    journal = "JHEP",
    volume = "07",
    pages = "020",
    year = "2025"
}

@article{Bar-Shalom:1998rqq,
    author = "Bar-Shalom, S. and Eilam, G. and Soni, A.",
    title = "{R-parity violation and CP violating and CP conserving spin asymmetries in lepton+ lepton- ---{\ensuremath{>}} sneutrino ---{\ensuremath{>}} tau+ tau-: Probing sneutrino mixing at LEP-2, NLC and mu mu colliders}",
    eprint = "hep-ph/9802251",
    archivePrefix = "arXiv",
    reportNumber = "UCRHEP-T-215, UCRHEP-T215",
    doi = "10.1103/PhysRevLett.80.4629",
    journal = "Phys. Rev. Lett.",
    volume = "80",
    pages = "4629--4632",
    year = "1998"
}

@article{Bullock:1992yt,
    author = "Bullock, B. K. and Hagiwara, Kaoru and Martin, Alan D.",
    title = "{Tau polarization and its correlations as a probe of new physics}",
    reportNumber = "KEK-TH-332, KEK-PREPRINT-92-33, DTP-92-36",
    doi = "10.1016/0550-3213(93)90045-Q",
    journal = "Nucl. Phys. B",
    volume = "395",
    pages = "499--533",
    year = "1993"
}

@article{Billur:2013rva, 
    author = "Billur, A. A. and Koksal, M.", 
    title = "{Probe of the electromagnetic moments of the tau lepton in gamma-gamma collisions at the CLIC}", 
    eprint = "1306.5620", 
    archivePrefix = "arXiv", 
    primaryClass = "hep-ph", 
    doi = "10.1103/PhysRevD.89.037301", 
    journal = "Phys. Rev. D", 
    volume = "89", 
    number = "3", 
    pages = "037301", 
    year = "2014"
}

@article{Rajaraman:2018uyb, 
    author = "Rajaraman, Arvind and Howard, Jessica N. and Riley, Rebecca and Tait, Tim M. P.", 
    title = "{The \(\tau\) Magnetic Dipole Moment at Future Lepton Colliders}",
    eprint = "1810.09570", 
    archivePrefix = "arXiv", 
    primaryClass = "hep-ph", 
    reportNumber = "UCI-HEP-TR-2018-11", 
    doi = "10.31526/lhep.2.2019.113", 
    journal = "LHEP", 
    volume = "2", 
    number = "2", 
    pages = "5", 
    year = "2019" 
}

@article{Koksal:2018xyi,
    author = {K{\"o}ksal, M.},
    title = "{Search for the electromagnetic moments of the $\tau$ lepton in photon{\textendash}photon collisions at the LHeC and the FCC-he}",
    eprint = "1809.01963",
    archivePrefix = "arXiv",
    primaryClass = "hep-ph",
    doi = "10.1088/1361-6471/ab0b53",
    journal = "J. Phys. G",
    volume = "46",
    pages = "065003",
    year = "2019"
}

@article{Wang:2024bfc,
    author = "Wang, ZeQiang",
    title = "{Probing $\tau$ lepton dipole moments at future Muon Colliders}",
    eprint = "2410.12663",
    archivePrefix = "arXiv",
    primaryClass = "hep-ph",
    reportNumber = "IRMP-CP3-24-30",
    doi = "10.22323/1.476.0327",
    journal = "PoS",
    volume = "ICHEP2024",
    pages = "327",
    year = "2025"
}

@article{Sun:2024vcd,
    author = "Sun, Xulei and Wu, Yongcheng and Zhou, Xiaorong",
    title = "{Sensitivity study of the tau lepton electric dipole moment at the Super Tau-Charm Facility*}",
    eprint = "2411.19469",
    archivePrefix = "arXiv",
    primaryClass = "hep-ex",
    doi = "10.1088/1674-1137/adf6e0",
    journal = "Chin. Phys.",
    volume = "49",
    number = "11",
    pages = "113001",
    year = "2025"
}

@article{Bruce:2018yzs,
    author = "Bruce, Roderik and others",
    title = "{New physics searches with heavy-ion collisions at the CERN Large Hadron Collider}",
    eprint = "1812.07688",
    archivePrefix = "arXiv",
    primaryClass = "hep-ph",
    doi = "10.1088/1361-6471/ab7ff7",
    journal = "J. Phys. G",
    volume = "47",
    number = "6",
    pages = "060501",
    year = "2020"
}

@article{dEnterria:2022sut,
    author = "d'Enterria, David and others",
    title = "{Opportunities for new physics searches with heavy ions at colliders}",
    eprint = "2203.05939",
    archivePrefix = "arXiv",
    primaryClass = "hep-ph",
    doi = "10.1088/1361-6471/acc197",
    journal = "J. Phys. G",
    volume = "50",
    number = "5",
    pages = "050501",
    year = "2023"
}

@article{Bernreuther:2004jv,
    author = "Bernreuther, W. and Brandenburg, A. and Si, Z. G. and Uwer, P.",
    title = "{Top quark pair production and decay at hadron colliders}",
    eprint = "hep-ph/0403035",
    archivePrefix = "arXiv",
    reportNumber = "CERN-PH-TH-2004-046, DESY-04-026, PITHA-04-06, TTP04-03",
    doi = "10.1016/j.nuclphysb.2004.04.019",
    journal = "Nucl. Phys. B",
    volume = "690",
    pages = "81--137",
    year = "2004"
}

@article{Lu:2025heu,
    author = "Lu, Peng-Cheng and Si, Zong-Guo and Zhang, Han",
    title = "{Probing the electromagnetic dipole moment of the {\ensuremath{\tau}} lepton in the e+e-{\textrightarrow}{\ensuremath{\gamma}}*/Z{\textrightarrow}{\ensuremath{\tau}}+{\ensuremath{\tau}}- reaction}",
    eprint = "2506.19557",
    archivePrefix = "arXiv",
    primaryClass = "hep-ph",
    doi = "10.1103/hvz6-b69m",
    journal = "Phys. Rev. D",
    volume = "112",
    number = "7",
    pages = "075039",
    year = "2025"
}

\end{document}